\documentclass[journal, onecolumn]{IEEEtran}
\usepackage{amsfonts}
\usepackage{amssymb}
\usepackage{stfloats}
\usepackage{cite}
\usepackage{graphicx}
\usepackage{psfrag}
\usepackage{subfigure}
\usepackage{amsmath}
\usepackage{array}
\usepackage{color}

\newtheorem{Thm}{Theorem}
\newtheorem{Lem}{Lemma}

\newtheorem{Def}{Definition}

\newtheorem{Ass}{Assumption}

\begin{document}

\title{Multi-Relay Selection Design and Analysis for Multi-Stream Cooperative Communications}

\author{Shunqing~Zhang,
        Vincent~K.~N.~Lau,~\IEEEmembership{Senior~Member,~IEEE}
\thanks{This paper is funded by RGC 615609. The results in this paper were
presented in part at the IEEE International Conference on
Communications, May 2008.}
\thanks{S.~Zhang was with the Department
of Electronic and Computer Engineering, Hong Kong University of
Science and Technology, Hong Kong. He is now with Huawei
Technologies, Co. Ltd., China (e-mail: sqzhang@huawei.com).}
\thanks{V.~K.~N.~Lau is with the Department
of Electronic and Computer Engineering, Hong Kong University of
Science and Technology, Hong Kong (e-mail: eeknlau@ust.hk).}}

\markboth{To appear in IEEE Transactions on Wireless
Communications,~2010} {{Zhang and Lau}: Multi-Relay Selection Design
and Analysis for Multi-Stream Cooperative Communications}

\maketitle

\begin{abstract}
In this paper, we consider the problem of multi-relay selection for
multi-stream cooperative MIMO systems with $M$ relay nodes.
Traditionally, relay selection approaches are primarily focused on
selecting one relay node to improve the transmission reliability
given a single-antenna destination node. As such, in the cooperative
phase whereby both the source and the selected relay nodes transmit
to the destination node, it is only feasible to exploit cooperative
spatial diversity (for example by means of distributed space time
coding). For wireless systems with a multi-antenna destination node,
in the cooperative phase it is possible to opportunistically
transmit multiple data streams to the destination node by utilizing
multiple relay nodes. Therefore, we propose a low overhead
multi-relay selection protocol to support multi-stream cooperative
communications. In addition, we derive the asymptotic performance
results at high SNR for the proposed scheme and discuss the
diversity-multiplexing tradeoff as well as the
throughput-reliability tradeoff. From these results, we show that
the proposed multi-stream cooperative communication scheme achieves
lower outage probability compared to existing baseline schemes.
\end{abstract}

\section{Introduction}
Cooperative communications for wireless systems has recently
attracted enormous attention. By utilizing cooperation among
different users, spatial diversity can be created and this is
referred to as \emph{cooperative diversity}
\cite{Laneman03,Sendonaris031,Nosratinia04}. In \cite{Laneman03},
the authors considered the case where multiple relay nodes are
available to assist the communication between the source and the
destination nodes using the decode-and-forward (DF) protocol, and
they showed that a diversity gain of $M(1-2r)$ can be achieved with
$M$ relay nodes and a multiplexing gain of $r$. However, the
advantages of utilizing more relay nodes is coupled with the
consumption of additional system resources and power, so it is
impractical (or even infeasible) to activate many relay nodes in
resource- or power-constrained systems. As a result, various relay
selection protocols have been considered in the literature. For
example, in \cite{Bletsas06,Molisch06,Azarian05,Kumar08} the authors
considered cooperative MIMO systems with a single-antenna
destination node and the selection of one relay node in the
cooperative phase. In \cite{Bletsas06} the authors proposed an
opportunistic relaying protocol and showed that this scheme can
achieve the same diversity-multiplexing tradeoff (DMT) as systems
that activate all the relay nodes to perform distributed space-time
coding. In \cite{Molisch06} the authors applied fountain code to
facilitate exploiting spatial diversity. In \cite{Azarian05,Kumar08}
the authors proposed a dynamic decode-and-forward (DDF) protocol
which allows the selected relay node to start transmitting as soon
as it successfully decodes the source message. On the other hand,
there are a number of works \cite{Yuksel07,Bolcskei06,Borade07} that
studied the capacity bounds and asymptotic performance (e.g. DMT
relation) for single-stream cooperative MIMO systems with a
single-antenna destination node. In \cite{Yuksel07} the authors
derived the DMT relation with multiple full-duplex relays and showed
that the DMT relation is the same as the DMT upper bound for
point-to-point MISO channels. In \cite{Bolcskei06} the authors
studied the network scaling law based on the amplify-and-forward
(AF) relaying protocol. In all the above works, the destination node
is assumed to have single receive antenna and hence, only one data
stream is involved in the cooperative phase. When the destination
node has multiple receive antennas, the system could support
multiple data streams in the cooperative phase and this could lead
to a higher spectral efficiency.

In this paper, we design a relay selection scheme for multi-stream
cooperative systems and analyze the resultant system performance. In
order to effectively implement multi-stream cooperative systems,
there are several technical challenges that require further
investigations.

\begin{list}{\labelitemi}{\leftmargin=0.5em}
\item {\bf How to select multiple relay nodes to support multi-stream cooperation in the cooperative phase?}
Most of the existing relay selection schemes are designed with
respect to having a single-antenna destination node and supporting
cooperative spatial diversity. For example, in \cite{Nikjah08} the
authors considered a rateless-coded system and proposed to select
relay nodes for supporting cooperative spatial diversity based on
the criterion of maximizing the received SNR.
In order to support multi-stream
cooperation, the relay selection metric should represent the
\emph{holistic} channel condition between all the selected relay
nodes and the destination node, but this property cannot be
addressed by the existing relay selection schemes.

\item {\bf How much additional benefit can multi-stream cooperation achieve?}
The spectral efficiency of the cooperative phase can be
substantially increased with multi-relay multi-stream cooperation
compared to conventional schemes. However, the system performance
may be bottlenecked by the source-relay links. Therefore, we
characterize the advantage of multi-stream cooperation in terms of
the end-to-end performance gain over conventional schemes that are
based on cooperative spatial diversity.
\end{list}

We propose a multi-stream cooperative relay protocol (DF-MSC-opt)
for cooperative systems with a multi-antenna destination node. We
consider optimized node selection in which a set of relay nodes is
selected for {\em multi-stream cooperation}. Based on the optimal
relay selection criterion, we compare the outage capacity, the
diversity-multiplexing tradeoff (DMT), and the
throughput-reliability tradeoff (TRT) of the proposed DF-MSC-opt
scheme against traditional reference baselines.

\emph{Notation}: In the sequel, we
adopt the following notations. $\mathbb{C}^{M \times N}$ denotes the
set of complex $M \times N$ matrices; $\mathbb{Z}$ denotes the set
of integers; upper and lower case letters denote matrices and
vectors, respectively; $( \cdot )^T$ denotes matrix transpose;
$\textrm{Tr}( \cdot )$ denotes matrix trace; $\textrm{diag}(x_1,
\ldots, x_L)$ is a diagonal matrix with entries $x_1, \ldots, x_L$;
$\mathcal{I} ( \cdot )$ denotes the indicator function; $\textrm{Pr}
( X )$ denotes the probability of event $X$; $\mathbf I_{N}$ denotes
the $N \times N$ identity matrix; $\lceil \cdot \rceil$ and $\lfloor
\cdot \rfloor$ denote the ceiling and floor operations,
respectively; $(\cdot)^{+} = \max(\cdot,0)$; $\doteq$, $\dot{\leq}$
and $\dot{\geq}$ denote exponential equality and inequalities where,
for example, $f(\rho) \doteq \rho^{b}$ if $\lim_{\rho \rightarrow
\infty} \frac{\log(f(\rho))} {\log(\rho)} = b$; $\mathbb{E}_{Y}[
\cdot ]$ denotes expectation over $Y$; and $F(x,k)$ denotes the
$\chi^2$ cumulative distribution function (CDF) for value $x$ and
degrees-of-freedom $k$.

\section{Relay Channel Model}
\label{sect:chan_mod} We consider a system consisting of a
single-antenna source node, $M$ half-duplex single-antenna relay
nodes, and a destination node with $N_r$ antennas.
For notational convenience, we denote
the source node as the $0^{\textrm{th}}$ node and the $M$ relay
nodes as the $\{1, 2, \ldots , M\}$-th node. We focus on block
fading channels such that the channel coefficients for all links
remain constant throughout the transmission of a source message.

We divide the transmission of a source message that requires $N$
channel uses into two phases, namely the \emph{listening phase} and
\emph{cooperative phase}. In the listening phase, all the relay
nodes listen to the signals transmitted by the source node until $K$
out of the $M$ relay nodes can decode the source
message\footnote{The relay system cannot enter the cooperative phase
if less than $K$ relay nodes can decode the source message within
$N$ channel uses.}. In the cooperative phase, the destination node
chooses $N_r$ nodes from amongst the source node and the
successfully decoding relay nodes to transmit multiple data streams
to the destination node. Specifically, let $\mathbf{x} =
[x(1),x(2),\ldots,x(N)]^{T}\in \mathbb{C}^{N \times 1}$ denote the
signals transmitted by the source node over $N$ channel uses.
Similarly, let $\mathbf{x}_m = [x_m(1), x_m(2), \ldots, x_m(N)]^{T}
\in \mathbb{C}^{N \times 1}$, $m=1,\ldots,M$, denote the signals
transmitted by the $m^{\textrm{th}}$ relay node over $N$ channel
uses. The signals received by the $m^{th}$ relay node is given by
$\mathbf y_{m} = [y_{m}(1), y_{m}(2),\ldots,y_{m}(N)]^{T} \in
\mathbb{C}^{N \times 1}$, $m=1,\ldots,M$, where
\begin{equation}
y_{m}(n) = h_{SR,m}x(n) + z_{m}(n),\label{eqn:eddyRxSigRelay}
\end{equation}
$h_{SR,m}$ is the fading channel coefficient between the source node
and $m^{th}$ relay node, $\textbf{H}_{SR} = [ h_{SR,1}, h_{SR,2},
\ldots, h_{SR,M} ]^T \in \mathbb{C}^{M \times 1}$ is a vector
containing the channel coefficients between the source and the $M$
relay nodes, and $z_{m}(n)$ is the additive noise with power
normalized to unity. Each relay node attempts to decode the source
message with each received signal observation until it can
successfully decode the message. The listening phase ends and the
cooperative phase begins after $K$ relay nodes successfully decodes
the source message. In the cooperative phase, let $\tilde x_k(n)$,
$k = 1, \ldots, K$, denote the signal relayed by the $k^{th}$
successfully decoding relay node and the aggregate signal
transmitted in the cooperative phase can be expressed as
$\mathbf x_D(n) = [x(n), \tilde
x_1(n), \tilde x_2(n), \ldots, \tilde x_K(n)]^{T} \in
\mathbb{C}^{(K+1) \times 1}$. Accordingly, the received signals at
the destination node are given by $\mathbf{Y} = [\mathbf y(1),
\mathbf y(2),\ldots,\mathbf y(N)]^{T} \in \mathbb{C}^{N \times
N_r}$, where
\begin{eqnarray*}
\label{eqn:rec_sig} \mathbf y(n) = \left\{
\begin{array}{l l}
\mathbf H_{SD}x(n) + \mathbf z(n) & \textrm{for the listening phase}\\
\mathbf H_{D}(\mathcal D) \mathbf V \mathbf x_D(n) + \mathbf z(n) &
\textrm{for the cooperative phase}
\end{array}
\right.
\end{eqnarray*}
$\mathbf H_{SD} \in \mathbb{C}^{N_r \times 1}$ represents the fading
channel coefficients between the source and the destination nodes,
$\mathcal D$ is the set representing the $K$ successfully decoding
relay nodes, $\mathbf H_{D}(\mathcal
D) = [\mathbf H_{SD}, \tilde{\mathbf H}_{RD,1}, \tilde{\mathbf
H}_{RD,2}, \ldots, \tilde{\mathbf H}_{RD,K}] \in \mathbb{C}^{N_r
\times (K+1)}$ represents the aggregate channel in the cooperative
phase with $\tilde{\mathbf H}_{RD,k} \in \mathbb{C}^{N_r \times 1}$
being the fading channel coefficients between $k^{th}$ successfully
decoding relay and the destination node, $\mathbf z(n) \in
\mathbb{C}^{N_r \times 1}$ is the additive noise with power
normalized to unity, and $\mathbf V$ is the \emph{node selection
matrix}. The selection matrix is
defined as $\mathbf V = \textrm{diag}(v_0, v_1,v_2,\ldots,v_{K})$,
where $v_{k} = 1$ if the $k^{\textrm{th}}$ node ($k \in \{ 0,
\ldots, K \}$) is selected to transmit in the cooperative phase and
$v_{k} = 0$ if the node is not selected.

The following assumptions are made throughout the rest of the paper.
\begin{Ass}[Half-duplex relay model] The half-duplex relay nodes can either transmit or
receive during a given time interval but not both.~
\hfill\IEEEQEDclosed
\end{Ass}
\begin{Ass}[Fading model] We assume block fading channels such that the channel
coefficients $\mathbf H_{SR}$ and $\mathbf H_{D} (\mathcal D)$
remain unchanged within a fading block (i.e., $N$ channel uses).
Moreover, we assume the fading channel coefficients of the
source-to-relay (S-R) links, relay-to-destination (R-D) links, and
source to destination (S-D) links are independent and identically distributed
(i.i.d.) complex symmetric random Gaussian variables with zero-mean
and variance $\sigma_{SR}^{2}$, $\sigma_{RD}^{2}$ and
$\sigma_{SD}^{2}$, respectively.~
\hfill\IEEEQEDclosed
\end{Ass}
\begin{Ass}[CSI model] \label{ass:CSI}Each relay node has perfect channel state information
(CSI) of the link between the source node and itself. The
destination node has perfect CSI of the S-D link and all R-D links.
For notational convenience, we denote the aggregate CSI as
$\mathbf{H} = \big( \mathbf H_{SR}, \mathbf H_{D} (\mathcal D)
\big)$.~ \hfill\IEEEQEDclosed
\end{Ass}
\begin{Ass}[Transmit power
constraints] The transmit power of
the source node is limited to $\rho_S$. The transmit power of the
$k^{th}$ relay node is limited to $\rho_k$.~ \hfill\IEEEQEDclosed
\end{Ass}

\section{Problem Formulation}
\label{sect:protocol} In this section, we first present the
encoding-decoding scheme and transmission protocol of the proposed
DF-MSC-opt scheme. Based on that, we formulate the multi-relay
selection problem as a combinatorial optimization problem.
\subsection{Encoding and Decoding
Scheme} \label{subsect:coding} The proposed multi-stream cooperation
system is facilitated by random coding and maximum-likelihood (ML)
decoding \cite{Cover91}. At the source node, an $R$-bit message W,
drawn from the index set $\{1, 2, \ldots , 2^R\}$, is encoded
through an encoding function $\mathcal{X}^N : \{ 1, 2, \ldots 2^R \}
\rightarrow \textbf{X}^N$. The encoding function at the source node
can be characterized by a {\em vector codebook} $\mathcal{C} = \{
\textbf{X}^N(1), \textbf{X}^N(2), \ldots, \textbf{X}^N(2^R) \} \in
\mathbb{C}^{N_r \times N}$. The $m^{\textrm{th}}$ codeword of
codebook $\mathcal{C}$ is defined as
\begin{IEEEeqnarray}{l}
\mathbf X^{N}(m) = \left[
\begin{array}{lll}
x_{1}^{(m)}(1) & \ldots & x_{1}^{(m)}(N_1)\\
& \vdots &\\
x_{N_r}^{(m)}(1) & \ldots & x_{N_r}^{(m)}(N_1)
\end{array}
\right.\left|
\begin{array}{lll}
x_{1}^{(m)}(N_1\!+\!1) & \ldots & x_{1}^{(m)}(N)\\
& \vdots &\\
x_{N_r}^{(m)}(N_1\!+\!1) & \ldots & x_{N_r}^{(m)}(N)
\end{array}\right],m=1,\ldots,2^R,\;\;\;\label{eqn:codebookEddy}
\end{IEEEeqnarray}
which consists of $N$ vector symbols of dimension $N_r \times 1$,
and $x_{k}^{(m)}(n)$ is the symbol to be transmitted by the
$k^{\textrm{th}}$ antenna during the $n^{\textrm{th}}$ channel use.
The vector codebook $\mathcal{C}$ is known to all the $M$ relay
nodes (for decoding and re-encoding) as well as the destination node
for decoding. At the receiver side (relay node or destination node),
the receiver decodes the R-bit message based on the observations,
the CSI, and a decoding function $\mathcal{Y}^N : (\textbf{Y}^N
\times \textbf{H}) \rightarrow \{ 1, 2, \ldots, 2^R \}$. We assume
ML detection in the decoding process. The detailed operation of the
source node and the relay nodes in the listening and cooperative
phases are elaborated in the next subsection.
\subsection{Transmission Protocol
for Multi-Stream Cooperation} \label{subsect:protdesign} 
\begin{figure}[h]
\centering
\includegraphics[width = 6in]{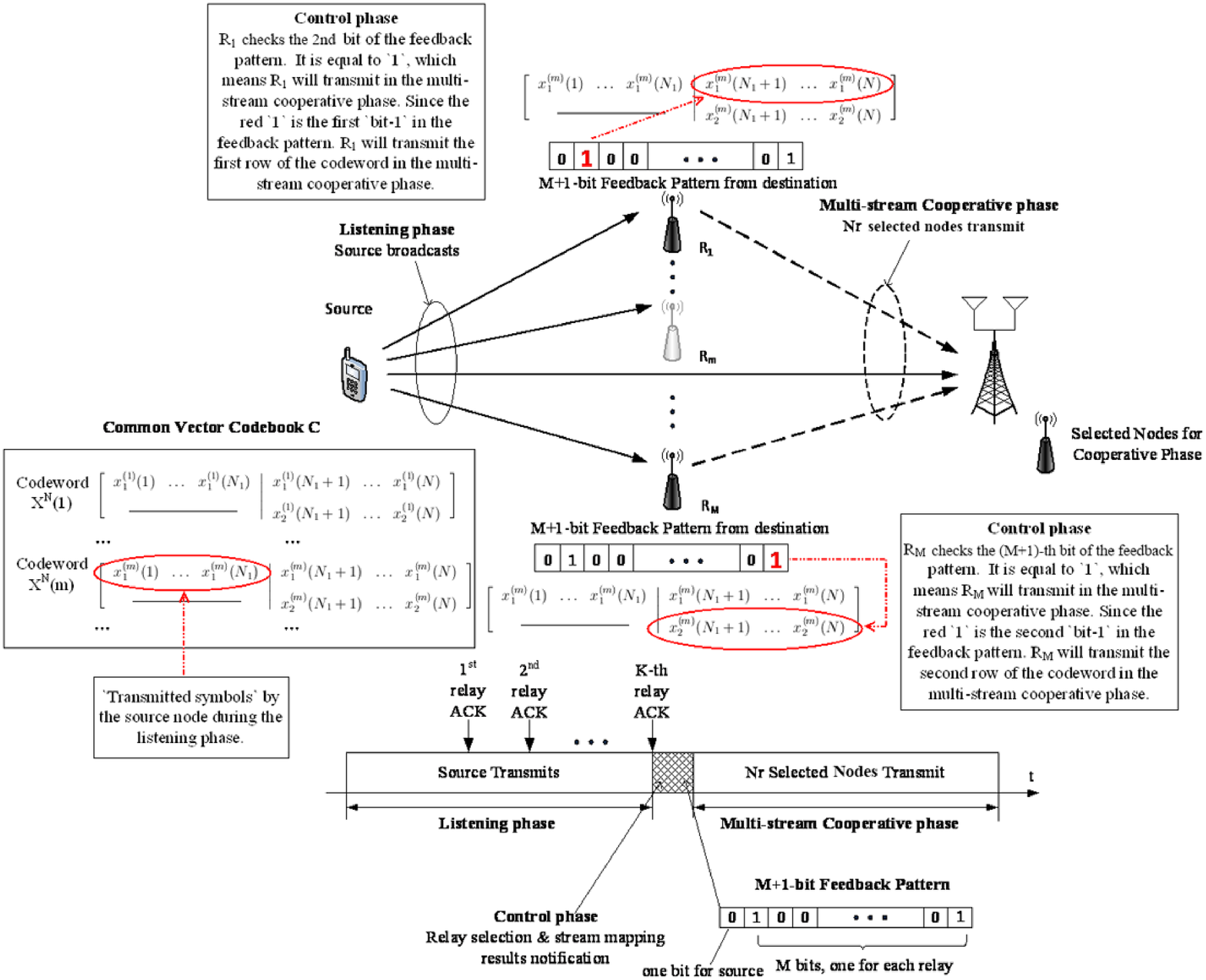}
\caption{Illustration of timing diagram and an example of feedback
pattern design. In this example, two relay nodes $R_1$ and $R_M$ are
selected to transmit in the cooperative phase. The selected relay
nodes will re-encode the same codeword (codeword $\mathbf X^{N}(m)$
in this example) selected from the same common vector codebook
$\mathcal C$ and the $R$-bit message received in the listening
phase. $R_1$ will send out the ``first'' row of codeword $\mathbf
X^{N}(m)$ and $R_M$ will send out the ``second'' row.}
\label{fig:fb_pat}
\end{figure}

\begin{figure}[h]
\centering
\includegraphics[width = 6in]{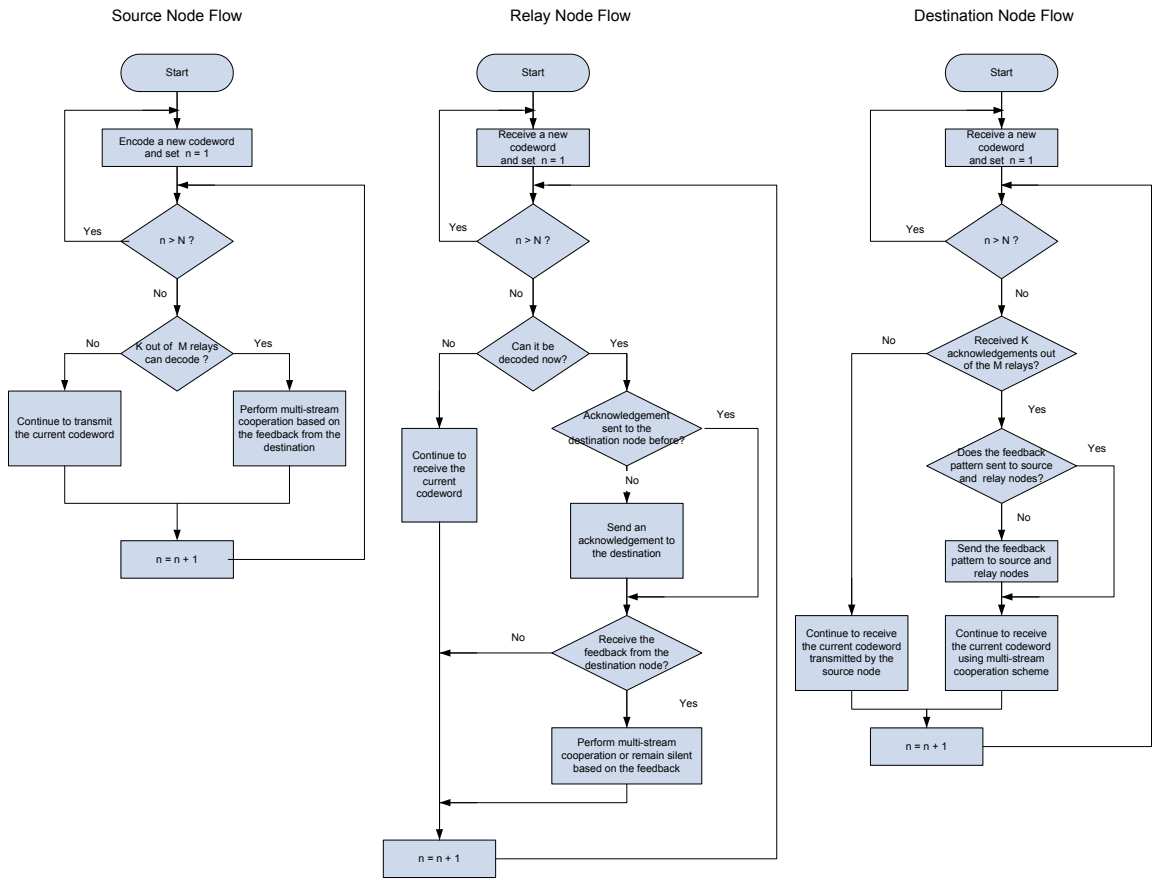}
\caption{Flow chart of the protocols at the source node, the relay node and the destination node.}
\label{fig:fc}
\end{figure}

The
proposed transmission protocol is illustrated in
Fig.~\ref{fig:fb_pat}, and the flow charts for the processing by the
source, relay, and destination nodes are shown in Fig.~\ref{fig:fc}.
Specifically, the $N$-symbol source message codeword (cf.
(\ref{eqn:codebookEddy})) is transmitted to the destination node
over two phases; the listening phase that spans the first $N_1$
channel uses and the cooperative phase that spans the remainder $N_2
= N - N_1$ channel uses.
\subsubsection{Listening phase} 
\begin{figure}[h]
\centering
\includegraphics[width = 3in]{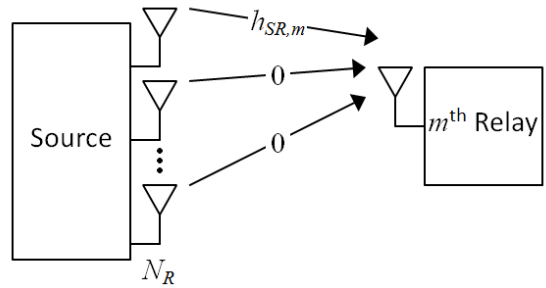}
\caption{The single-antenna
source-relay channel is equivalent to a multi-antenna virtual MISO
channel.} \label{fig:virtualMIMO}
\end{figure}

In
the listening phase, the single-antenna source node transmits the
first row\footnote{The source node has a single transmit antenna and
hence, could only transmit one row of the vector codeword $X^N(m)$
during the listening phase.} of the message codeword
(i.e. $[ x_{1}^{(m)}(1) \ldots
x_{1}^{(m)}(N_1) ]$ as per (\ref{eqn:codebookEddy})) to the relay
and destination nodes. Each relay node attempts to decode the source
message with each received signal observation. Although the source
node transmits only the first row of the source message codeword,
the relay nodes can still detect the source message using standard
random codebook and ML decoding argument.
Effectively, we can visualize a {\em
virtual system with a multi-antenna source node} as shown in
Fig.~\ref{fig:virtualMIMO}, and the {\em missing rows} in the
message codeword transmitted by the source node is equivalent to
channel erasure in a virtual MISO source-relay channel. Once a
relay node successfully decodes the source message, it sends an
acknowledgement $ACK_{RD}$ to the destination node through a
dedicated zero-delay error free feedback link.
\subsubsection{Control phase and
signaling scheme}\label{sect:controlPhase} Without loss of
generality, we assume that $K$ relay nodes can successfully decode
the source message with $N_1$ received signal observations. Upon
receiving the acknowledgement from the $K$ successfully decoding
relay nodes, the destination node enters the \emph{control} phase
and selects $N_r$ nodes to participate in the multi-stream
cooperation phase (cf. Section~\ref{subsect:probfor}). Specifically,
the destination node indicates to the source and relay nodes the
transition to the cooperative phase as well as the node selection
decisions via an $(M+1)$-bit feedback pattern.
The first bit of the feedback
pattern is used to index the $0^{\textrm{th}}$ node (the source
node) and the last $M$ bits are used to index the $M$ relay nodes.
The feedback pattern contains $N_r$
bits that are set to 1; the $m^{\textrm{th}}$ bit of the feedback
pattern is set to 1 if the corresponding node is selected to
participate in the cooperative phase, whereas the bit is set to 0 if
the node is not selected. Note that the total number of feedback
bits required by the proposed multi-stream cooperation scheme is $K$
$ACK_{RD}$ plus one feedback
pattern with $M+1$ bits, which is less than 2 bits per relay node.
\subsubsection{Cooperative phase} In
the cooperative phase, the $N_r$ selected nodes cooperate to
transmit the \hbox{$(N_1+1)^{\textrm{th}}$} to the
\hbox{$N^{\textrm{th}}$} columns of the source message codeword (cf.
(\ref{eqn:codebookEddy})) to the destination node to assist it with
decoding the source message. Specifically, for $k = 1, \ldots, N_r$,
the node corresponding to the $k^{th}$ bit that is set to 1 in
the feedback pattern transmits the $k^{\textrm{th}}$ row of the
message codeword (i.e. $[
x_{k}^{(m)}(N_1+1) \ldots x_{k}^{(m)}(N) ]$ as per
(\ref{eqn:codebookEddy})) to the destination node.

To better illustrate the proposed
transmission protocol, we show in Fig.~\ref{fig:fb_pat} an example
of the proposed system with a destination node with $N_r = 2$
antennas. Suppose the feedback pattern is given as follows:
\begin{center}
  \begin{tabular}{ r | c | c | c | c | c | c | c | }
  \multicolumn{1}{r}{}
  & \multicolumn{1}{c}{$\!\!\!\!\substack{\textrm{source}\\\textrm{node}}\!\!\!\!$}
  & \multicolumn{6}{c}{$|\!\!\leftarrow\;\;\;\substack{\textrm{$M$ relay nodes}}\;\;\;\rightarrow\!\!|$}\\
  \cline{2-8}
  Feedback Pattern & 0 & 1 & 0 & 0 & $\cdots$ & 0 & 1 \\
  \cline{2-8}
  \multicolumn{1}{r}{}
  & \multicolumn{7}{c}{$|\!\!\leftarrow\;\;\;\;\;\;\;\;\substack{\textrm{$M+1$ bits}}\;\;\;\;\;\;\;\;\rightarrow\!\!|$}\\
  \end{tabular}
\end{center}
This corresponds to selecting the first relay node ($R_1$) and the
$M^{\textrm{th}}$ relay node ($R_M$) to participate in the
cooperative phase. In the listening phase, the source node transmits
the first row of the codeword $X^N(m)$ given by $[ x_{1}^{(m)}(1)
\ldots x_{1}^{(m)}(N_1) ]$. In the cooperative phase, $R_1$ and
$R_M$ transmit $[ x_{1}^{(m)}(N_1+1) \ldots x_{1}^{(m)}(N) ]$ and $[
x_{2}^{(m)}(N_1+1) \ldots x_{2}^{(m)}(N) ]$, respectively.
Therefore, the effective transmitted codeword can be expressed as
\begin{IEEEeqnarray*}{l}
\mathbf X^{N}(m) = \Bigg[ \overbrace{\begin{array}{ccc}
x_{1}^{(m)}(1) & \ldots & x_{1}^{(m)}(N_1)\\
\multicolumn{3}{c}{\rule[0.2cm]{3cm}{0.01in}}
\end{array}}^{\substack{\textrm{Transmitted by source}\\\textrm{node in listening phase}}}
\Bigg| \underbrace{\overbrace{\begin{array}{lll}
x_{1}^{(m)}(N_1+1) & \ldots & x_{1}^{(m)}(N)\\
x_{2}^{(m)}(N_1+1) & \ldots & x_{2}^{(m)}(N)
\end{array}}^{\substack{\textrm{Transmitted by $R_1$}\\\textrm{in cooperative phase}}}}_{\substack{\textrm{Transmitted by $R_M$}\\\textrm{in cooperative phase}}}\Bigg].
\end{IEEEeqnarray*}

\subsection{Problem Formulation for Multi-stream Cooperation}
\label{subsect:probfor} We focus on node selection by the
destination node for optimizing outage performance (cf.
Section~\ref{sect:controlPhase}), where the destination node only
has CSI of the S-D and R-D links $\mathbf H_{D}(\mathcal D)$ (cf.
Assumption~\ref{ass:CSI}). The proposed Multi-stream Cooperation Scheme with Optimized node
selection is called DF-MSC-opt and we define the outage event as
follows:
\begin{Def} [Outage]\label{def:outage}
Outage refers to the event when the total instantaneous mutual
information between the source and the destination nodes is less
than the target transmission rate $R$. Mathematically, outage can be
expressed as
\begin{equation}
\label{eqn:def_1} \mathcal{I}(I_{\textrm{DF-MSC-opt}}(\mathbf{H},
\mathbf V)<R),
\end{equation}
where $\mathbf H =(\mathbf H_{SR}, \mathbf H_{D} (\mathcal D))$ is
the aggregate channel realization and $\mathbf V$ is
the node selection action given
$\mathbf H_{D}(\mathcal D)$.~ \hfill\IEEEQEDclosed
\label{Def:outage}
\end{Def}

The total instantaneous mutual information of the multi-stream
cooperative system in (\ref{eqn:def_1}) is given by the following
theorem.
\begin{Thm} {\em (Instantaneous Mutual Information of Multi-Stream
Cooperative System)} \label{Thm:1} Given the aggregate
channel realization $\mathbf H =(\mathbf H_{SR}, \mathbf H_{D}
(\mathcal D))$, the instantaneous mutual information
$I_{\textrm{DF-MSC-opt}}(\mathbf{H},\mathbf V)$ (bits/second/channel
use) between the source and destination nodes for multi-stream
cooperative system can be expressed as
\begin{IEEEeqnarray}{l}
I_{\textrm{DF-MSC-opt}}(\mathbf{H},\mathbf
V)\!\!=\!\!\textstyle\frac{1}{N}\big\{N_1 \log(1\!+\!\rho_S |\mathbf
H_{SD}|^2)\!+\!N_2 \log \det(\mathbf I_{N_r}\!+\!\mathbf
H_{D}(\mathcal D) \mathbf V \mathbf \Gamma \mathbf V^{H} \mathbf
H_{D}(\mathcal D)^{H})\big\},\;\;\;\;\label{eqn:mul_info_1}
\end{IEEEeqnarray}
where $N_1$ and $N_2$ are the numbers of channel uses of the
listening phase and the cooperative
phase, respectively, and
$\mathbf \Gamma = \textrm{diag}
(\rho_S, \rho_1, \rho_2, \ldots, \rho_M)$ is the transmit power
matrix of the source node and the $M$ relay nodes. Note that $N_1$
and $N_2$ are random variables that depend on the realization of the
S-R links $\mathbf H_{SR}$.~ \hfill\IEEEQEDclosed
\end{Thm}

The proof of Theorem \ref{Thm:1} can
be extended from \cite{Cover91,Khojastepour03} by applying random
Gaussian codebook argument. Note that the first term in the mutual
information in (\ref{eqn:mul_info_1}) corresponds to the
contribution from the source
transmission\footnote{The mutual
information contributed by the source node can also be obtained from
the mutual information of the virtual multi-antenna source model in
Fig.~\ref{fig:virtualMIMO}.}. By virtue of the proposed DF-MSC-opt
scheme, multiple data streams are cooperatively transmitted in the
cooperative phase (unlike traditional schemes wherein only a single
stream is transmitted). Hence, we can fully exploit the spatial
channels created from the multi-antenna destination node and
therefore achieving higher mutual information in the second term of
(\ref{eqn:mul_info_1}).

By Definition~\ref{def:outage}, the average outage probability is
given by
\begin{equation*}
\mathcal P_{out}(\mathbf V) = \mathbb E_{\mathbf H}\big[\mathcal{I}
\big(I_{\textrm{DF-MSC-opt}}(\mathbf{H},\mathbf V) < R \big)\big],
\end{equation*}
which is a function of the node selection policy $\mathbf V$. It
follows that, given the channel realization $\textbf{H}$, the
optimal node selection policy is given by $\mathbf
{V}^\star\!=\!\displaystyle\arg\min_{\mathbf {V} \in
\mathbf{\Omega}} \mathcal{I} \big(I_{\textrm{DF-MSC-opt}}
(\mathbf{H},\mathbf V)\!<\!R \big)$, where the set of all feasible
node selection actions is defined as
\begin{eqnarray}
\mathbf{\Omega} = \{ \mathbf{\Lambda} \in
\{0,1\}^{(M+1)\times(M+1)}|\mathbf{\Lambda} \textrm{ is diagonal and
} \textrm{Tr}(\mathbf{\Lambda}) = Nr \}.\label{eqn:searchSpaceEddy}
\end{eqnarray}
Equivalently, for any transmission rate $R$, the optimal node
selection policy is given by
\begin{eqnarray}
\label{eqn:min_out_3} \mathbf {V}^\star & = & \arg\max_{\mathbf
{V}\in \mathbf \Omega}I_{\textrm{DF-MSC-opt}}(\mathbf H, \mathbf
V).\label{eqn:optimizationProblemEddy}
\end{eqnarray}

\section{Outage Analysis for the DF-MSC-opt Scheme}
\label{sect:performance} In this section, we derive the asymptotic
outage probability of the DF-MSC-opt scheme. For simplicity, we
assume $\sigma_{SD}^2 = \sigma_{RD}^2 = \sigma_{D}^2$ and the power
constraints of all the nodes are scaled up in the same
order\footnote{As such, there is no loss of generality to assume
$\Gamma = \rho_S \mathbf I_{M+1}$ when studying high SNR analysis.},
i.e., $\lim_{\rho_S \rightarrow
\infty} \frac{\rho_m}{\rho_S} = 1$ for all $m \in \{1,\ldots,M\}$.
The outage probability can be expressed as
\begin{eqnarray}
\label{eqn:outage_2}\mathcal {P}_{out}^\star = \mathbb{E}_{\mathbf
H}\big[\mathcal{I} \big(I_{\textrm{DF-MSC-opt}}(\mathbf{H},\mathbf
V^\star) < R \big)\big] \stackrel{(a)}{=} \mathbb{E}_{\mathbf
H_{SR}} \big[ \textrm{Pr} \big( I_{\textrm{DF-MSC-opt}}
(\mathbf{H},\mathbf V^\star) < R | \mathbf H_{SR} \big) \big],
\end{eqnarray}
where in (a) the randomness of
$I_{\textrm{DF-MSC-opt}}(\mathbf{H},\mathbf V^\star)$ is introduced
by the aggregate channel gain in the cooperative phase $\mathbf
H_{D} (\mathcal D)$ and the optimal node selection policy $\mathbf
V^\star$. Let $\mathcal H_{N_1}$ denote the collection of all
realizations of the S-R links $\mathbf H_{SR}$ such that $K$ out of
$M$ relay nodes can successfully decode the source message, and the
listening phase ends at the $(N_1)^{\textrm{th}}$ channel use. We
define the outage probability given $\mathbf H_{SR}$ and the
duration of the listening phase $N_1$ as $P_{SR}(\mathbf H_{SR},
N_1) = \textrm{Pr}\big( I_{\textrm{DF-MSC-opt}} (\mathbf{H},\mathbf
V^\star) < R | \mathbf H_{SR} \in \mathcal H_{N_1} \big),$
which includes the following cases.
\begin{list}{\labelitemi}{\leftmargin=0.5em}
\item{\emph{Case 1:}} \emph{The listening phase ends within $N$ channel uses.} In this case, $N_1 <
N$, $N_2 = N - N_1$, and the outage probability $P_{SR}(\mathbf
H_{SR}, N_1)$ can be expressed as
\begin{eqnarray}
\label{eqn:outage_4}\textstyle P_{SR}^{\textrm{Case 1}}(\mathbf
H_{SR}, N_1) = \textrm{Pr}\big[\big( \frac{N_1}{N} \log(1 + \rho_S
|\mathbf H_{SD}|^2) + \frac{N-N_1}{N} g(\mathbf H, \mathbf V^\star)
\big) < R\big],
\end{eqnarray}
where $g(\mathbf H, \mathbf V^\star) = \displaystyle\max_{\mathbf V
\in \mathbf \Omega}\textstyle\log\det\big(\mathbf I_{N_r} + \rho_S
\mathbf H_{D}(\mathcal D) \mathbf V \mathbf V^{H} \mathbf
H_{D}(\mathcal D)^{H}\big)$ for $K \geq N_r$, and $g(\mathbf H,
\mathbf V^\star) = \log\det\big(\mathbf I_{N_r} + \rho_S \mathbf
H_{D}(\mathcal D) \mathbf H_{D}(\mathcal D)^{H}\big)$ for $K < N_r$.

\item{\emph{Case 2:}} \emph{The listening phase lasts for more than $N$ channel
uses.} In this case, the outage event is only contributed by the
direct transmission between the source and the destination nodes.
The outage probability $P_{SR}(\mathbf H_{SR}, N_1)$ can be
expressed as
\begin{eqnarray}
\label{eqn:outage_5} P_{SR}^{\textrm{Case 2}}(\mathbf H_{SR}, N_1) =
\textrm{Pr}\big(\log(1 + \rho_S |\mathbf H_{SD}|^2) < R\big).
\end{eqnarray}
\end{list}

\begin{figure}[h]
\centering
\includegraphics[width = 3.5in]{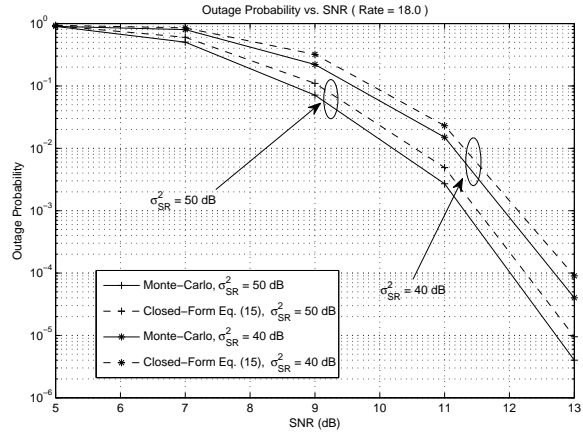}
\caption{Outage probability vs. SNR of the DF-MSC-opt scheme for
different $\sigma_{SR}^2$ under $N_r = 3$, $K =6$ and $M =15$.}
\label{fig:cf_1}
\end{figure}

\begin{figure}[h]
\centering
\includegraphics[width = 3.5in]{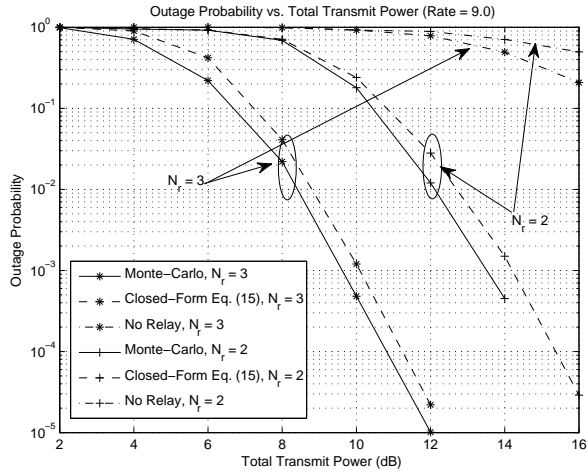}
\caption{Outage probability vs. SNR of the DF-MSC-opt scheme for
different $N_r$ under $\sigma_{SR}^2 = 10 dB$, $K=6$ and $M =15$.}
\label{fig:cf_2}
\end{figure}

Substituting \eqref{eqn:outage_4} and
\eqref{eqn:outage_5} into \eqref{eqn:outage_2}, the outage probability is given by
\begin{IEEEeqnarray}{l}
\label{eqn:outage_6}\textstyle\mathcal {P}_{out}^\star = \sum_{l = 1}^{N}P_{SR}^{\textrm{Case 1}}(\mathbf H_{SR}, l) \textrm{Pr}(\mathbf H_{SR} \in \mathcal H_{l}) + \sum_{l > N}P_{SR}^{\textrm{Case 2}}(\mathbf H_{SR}, l) \textrm{Pr}(\mathbf H_{SR} \in \mathcal H_{l}) \\
\;\;\;\;\;= \textstyle\sum_{l = 1}^{N}P_{SR}^{\textrm{Case
1}}(\mathbf H_{SR}, l) \textrm{Pr}(\mathbf H_{SR}\!\in\!\mathcal
H_{l})\!+\! \textrm{Pr}\big(\log(1\!+\!\rho_S|\mathbf
H_{SD}|^2)\!<\!R\big) \textrm{Pr}\left(\mathbf H_{SR}\!\in\!\cup_{l
> N} \mathcal H_{l}\right), \nonumber
\end{IEEEeqnarray}
where $\textrm{Pr}(\mathbf H_{SR} \in \mathcal H_{l})$ is the
probability that $K$ out of $M$ relay nodes can successfully decode
the message at the $l^{th}$ channel use. To evaluate the outage
probability, we calculate each term
in equation \eqref{eqn:outage_6} as follows. First, it can be shown
that
\begin{eqnarray}
\label{eqn:outage_7} \textrm{Pr}(\mathbf H_{SR} \in \mathcal
H_{N_1}) = \textrm{Pr}\left(\mathbf H_{SR} \in \cup_{l > N_1 - 1}
\mathcal H_{l}\right) -  \textrm{Pr}\left(\mathbf H_{SR} \in \cup_{l
> N_1} \mathcal H_{l}\right) = \Phi_{N_1-1} - \Phi_{N_1}
\end{eqnarray}
where
\begin{eqnarray}
\textstyle\Phi_{l} = \sum_{i = 0}^{K - 1}{M \choose i} \big(1 -
\exp(-\frac{2^{NR/l} - 1}{\rho_S \sigma_{SR}^2})\big)^{M-i}
\big(\exp(-\frac{2^{NR/l} - 1}{\rho_S \sigma_{SR}^2})\big)^{i},\;l =
1, \ldots, N,
\end{eqnarray}
denotes the probability that less than $K$ relay nodes can
successfully decode the source message at the $l^{th}$ channel use.
Second, $P_{SR}^{\textrm{Case 1}}(\mathbf H_{SR}, l)$ can be
upper-bounded\footnote{It is non-trivial to evaluate
$P_{SR}^{\textrm{Case 1}}(\mathbf H_{SR}, l)$ exactly due to the
dynamics of the optimal nodes selection.} as shown in the following
lemma.
\begin{Lem}
\label{Lem:exp} The outage probability for the DF-MSC-opt scheme
given that $K$ relay nodes can successfully decode the source
message at the $l^{\textrm{th}}$ channel use can be upper bounded as
\begin{eqnarray}
P_{SR}^{\textrm{Case 1}}(\mathbf H_{SR}, l) \leq \textrm{Pr}\left(
f(\alpha_l,\mathbf H, \mathbf V^\star) < R \right),\;l = 1, \ldots,
N,\label{eqn:outage_13}
\end{eqnarray}
where $\alpha_l = l/N$, $f(\alpha_l,\mathbf H, \mathbf V^\star) =
\alpha_l \log(1 + \rho_S |\mathbf H_{SD}|^2) + (1 - \alpha_l)\sum_{i
= 1}^{L_T} \log_2\big(1 + \rho_S \sigma_{D}^2 \kappa(i) \big)$ with
$L_T = \min(N_r,K+1)$ denoting the number of transmitted streams in
the cooperative phase, and $\kappa(i)$, $i=1,\ldots,L_T$, are ($L_T$
out of $K+1$) ordered $\chi^2$-distributed variables with $2 N_r$
degrees of freedom.
\end{Lem}
\proof Refer to Appendix \ref{pf:Lem_exp} for the proof.
\endproof
Third, when fewer than $K$ relay nodes can successfully decode the
source message within $N$ channel uses, the source node transmits to
the destination node with the direct link only; it follows that
$\textrm{Pr}(\mathbf H_{SR} \in \cup_{l
> N} \mathcal H_{l}) = \Phi_{N}$ and $\textrm{Pr}\big(\log(1 +
\rho_S |\mathbf H_{SD}|^2) < R\big) = F\left(\frac{2^{R} - 1
}{\rho_S \sigma_{D}^2};N_r\right)$. Therefore, the outage
probability (cf. \eqref{eqn:outage_6}) is given by
\begin{eqnarray}
\label{eqn:outage_14} \mathcal {P}_{out}^\star & = & \textstyle\Phi
F\left(\frac{2^{R} - 1 }{\rho_S \sigma_{D}^2};N_r\right) +
\sum_{\alpha_l = 1/N}^{1}(\Phi_{l-1} - \Phi_{l}) P_{SR}^{\textrm{Case 1}}(\mathbf H_{SR}, l) \nonumber \\
& \leq & \textstyle\Phi F\left(\frac{2^{R} - 1 }{\rho_S
\sigma_{D}^2};N_r\right) + \sum_{\alpha_l = 1/N}^{1}(\Phi_{l-1} -
\Phi_{l}) \textrm{Pr}\left( f(\alpha_l,\mathbf H, \mathbf V^\star) <
R\right) \nonumber
\\
& = & \textstyle\Phi F\left(\frac{2^{R} - 1 }{\rho_S
\sigma_{D}^2};N_r\right) - \Phi_{l'}' \textrm{Pr}\left(
f(\alpha_l',\mathbf H, \mathbf V^\star) < R\right),
\end{eqnarray}
where $\Phi_{l}'$ and $\Phi_{l'}'$ denote the first order derivative
of $\Phi_{l}$ with respect to $\alpha_l$ and evaluated at $l =
\alpha_l N$ and $l' = \alpha_l' N$, respectively. Note that in the last step of
\eqref{eqn:outage_14} we apply the {\em Mean Value Theorem}
\cite{Rudin76} which guarantees the existence of the point
$\alpha_l' \in (0,1)$. Moreover, the outage probability upper bound
of the DF-MSC-opt scheme is given by the following theorem.
\begin{Thm}
\label{Thm:2} For sufficiently large $N$, the outage probability
upper bound of the DF-MSC-opt scheme is given by
\begin{eqnarray}
\label{eqn:outage_16} \textstyle\mathcal {P}_{out}^\star \leq
\underbrace{\textstyle\Phi F\left(\frac{2^{R} - 1 }{\rho_S
\sigma_{D}^2};N_r\right)}_{\textrm{Direct Transmission}} -
\underbrace{\textstyle\Phi_{l'}'
F\bigg(\frac{2^{\frac{R}{\alpha_l'}} - 1 }{\rho_S
\sigma_{D}^2};N_r\bigg) F\bigg(\frac{2^{\frac{R}{1-\alpha_l'}} - 1
}{\rho_S \sigma_{D}^2 };N_r\bigg)^{K + 1}}_{\textrm{Relay-Assisted
Transmission}}.
\end{eqnarray}
\end{Thm}
\proof Refer to Appendix \ref{pf:Thm:2} for the proof.
\endproof
For systems without relays, the outage probability is given by
$\mathcal P_{out,pp} = F\left(\frac{2^{R} - 1 } {N_r \rho_S
\sigma_{D}^2};N_r\right)$. Note that the DF-MSC-opt scheme can
achieve lower outage probability by
taking advantage of multi-stream
cooperative transmission. We can interpret $-\Phi_{l'}'$ as the {\em
cooperative level}, which quantifies the probability that the relay
nodes can \emph{assist} the direct transmission. For example,
$-\Phi_{l'}' = 1$ implies that the DF-MSC-opt scheme can be fully
utilized, whereas $-\Phi_{l'}' = 0$ implies that the source node
transmits to the destination node with the direct link only. As per
\eqref{eqn:outage_14}, the outage probability $\mathcal
{P}_{out}^\star$ is a decreasing function with respect
to\footnote{For a sufficiently
large SNR $\rho_S$, $F\bigg(\frac{2^{\frac{R}{\alpha_l'}} - 1
}{\rho_S \sigma_{D}^2};N_r\bigg) F\bigg(\frac{2^{\frac{R}{L_T
(1-\alpha_l') }} - 1 }{\rho_S \sigma_{D}^2 };N_r\bigg)^{K + 1}$ is
much smaller than $F\left(\frac{2^{R} - 1 } {N_r \rho_S
\sigma_{D}^2};N_r\right)$.}$-\Phi_{l'}'$, whereas $-\Phi_{l'}'$ is a
decreasing function with respect to the strength of the S-R links
$\sigma_{SR}^2$. As shown in Fig. \ref{fig:cf_1}, when the strength
of the S-R links increases from 30dB to 40dB, the cooperative level
increases and hence the outage probability decreases. On the other
hand, as shown in \eqref{eqn:outage_16} and as illustrated in
Fig.~\ref{fig:cf_2}, the outage probability decreases with
increasing number of receive
antennas at the destination node.

\section{DMT and TRT Analyses for the DF-MSC-opt Scheme}
\label{sect:dmt} 
\begin{figure}[h]
\centering
\includegraphics[width = 3.5in]{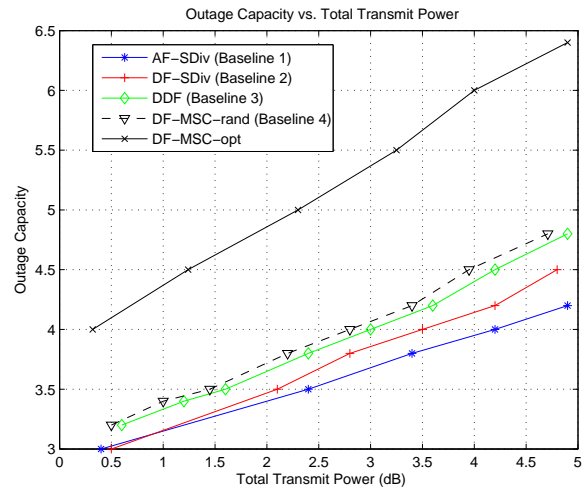}
\caption{Outage capacity comparison of different schemes for $N_r =
3$, $K = 3$ and $M = 15$. The channel variances of the S-R, R-D, S-D
links are normalized to unity.} \label{fig:out}
\end{figure}

In this section, we focus on characterizing the
performance of the DF-MSC-opt scheme in the high SNR regime.
Specifically, we perform DMT and TRT analyses based on the outage
probability $\mathcal {P}_{out}^\star$ as shown in
\eqref{eqn:outage_14}. There are two
fundamental reasons for the performance advantage of the proposed
DF-MSC-opt scheme, namely multi-stream cooperation and optimal node
selection $\textbf{V}^\star$ (cf.
\eqref{eqn:optimizationProblemEddy}). To illustrate the contribution
of the first factor, we shall compare the performance of DF-MSC-opt
with the following baselines:
\begin{list}{\labelitemi}{\leftmargin=0.5em}
\item \emph{DF-SDiv (Baseline 1):} \emph{The DF relay protocol for
cooperative spatial diversity.} The listening phase and the cooperative phase have fixed durations, i.e., each phase consists of $N/2$ channel uses. The source node and \emph{all} the
successfully decoding relay nodes cooperate using distributed
space-time coding. At the destination node, maximum ratio combining
(MRC) is used to combine the observations from the different receive
antennas.
\item \emph{AF-SDiv (Baseline 2):} \emph{The AF relay protocol for
cooperative spatial diversity.} All the relay nodes transmit a
scaled version of their soft observations in the cooperative phase.
At the destination node, MRC is used to combine the observations
from different receive antennas.
\item \emph{DDF (Baseline 3):} The \emph{dynamic DF protocol \cite{Azarian05}.}
Once a relay node successfully decodes the source message, it
immediately joins the transmission using distributed space-time
coding. At the destination node, MRC is used to combine the
observations from different receive antennas.
\end{list}
On the other hand, to illustrate the contribution of optimal node
selection, we shall compare the performance of the DF-MSC-opt scheme
with the DF-MSC-rand scheme (Baseline 4), which corresponds to a
similar multi-stream cooperation scheme but with randomized node
selection $\textbf{V}$ randomly generated from the node selection
space $\mathbf{\Omega}$ (cf. \eqref{eqn:searchSpaceEddy}).
Table~\ref{table:DF-MSCvsBaselines} summarizes the major differences
among the DF-MSC-opt scheme and the baseline schemes. We illustrate
in Fig.~\ref{fig:out} the outage capacity versus SNR (with
$\mathcal{P}_{out} = 0.01$) of the cooperative systems with $N_r =
3$, $K = 3$, $M = 15$. Note that the proposed DF-MSC-opt scheme can
achieve a gain of more than $1$ bit/channel use over the four
baseline schemes.

\subsection{DMT Analysis}
In order to analyze the DMT relation of the DF-MSC-opt scheme, we
first derive the relation between the outage probability and the
multiplexing gain. In the outage probability expression
\eqref{eqn:outage_14}, for a sufficiently large $\rho_S$, the asymptotic expression of the first
term is given by
\begin{eqnarray}
\textstyle\Phi F\left(\frac{2^{R} - 1 } {\rho_S
\sigma_{D}^2};N_r\right) \dot{=}
\sum_{i = 0}^{K - 1}{M \choose i} \rho_S^{-(M-i)(1 -
r)^{+}}\rho_S^{-N_r(1 - r)^{+}} \doteq \rho_S^{-(M - K + 1 + N_r)(1
- r)^{+}}, \label{eqn:outage_22}
\end{eqnarray}
and the asymptotic expression of $\Phi'_{l'}$ is given by
\begin{eqnarray}
\Phi'_{l'} & = & \textstyle\sum_{i = 0}^{K - 1}{M \choose i} (1 -
\phi_{l'})^{M - i - 1} (\phi_{l'})^{i - 1} ( (M-i)(1 - \phi_{l'})'
\phi_{l'} + i (1
- \phi_{l'}) \left(\phi_{l'}\right)' ) \nonumber\\
& \dot{=} & \textstyle\sum_{i =
0}^{K - 1} {M \choose i} \rho_S^{-(M - i)\big(1 -
\frac{r}{\alpha_l'}\big)^{+}} \dot{\leq} \rho_S^{-\min_{i}(M -
i)\big(1 - \frac{r}{\alpha_l'} \big)^{+}} = \rho_S^{-(M - K +
1)\big(1 - \frac{r}{\alpha_l'}\big)^{+}}\label{eqn:outage_23}
\end{eqnarray}
where $\phi_{l'} =
\exp\Big(-\frac{2^{\frac{r\log\rho_S}{\alpha_{l'}} - 1}}{\rho_S
\sigma_{SR}^2}\Big)$. Moreover, we can express the term
$\textrm{Pr}\left(f(\alpha_l',\mathbf H,\mathbf V) < R\right)$ in
\eqref{eqn:outage_14} as
\begin{IEEEeqnarray}{l}
\textstyle\textrm{Pr}\left(f(\alpha_l',\mathbf H,\mathbf V) <
R\right) \doteq \textstyle\textrm{Pr} \big( \rho_S^{\alpha_l'(1 -
\delta)^{+}} \prod_{i=1}^{L_T}\rho_S^{(1-\alpha_l')(1 -
\beta_i)^{+}} < \rho_S^{r}\big) \nonumber \\
\;\;\;\;\;\;\;\;\;\;= \textstyle\textrm{Pr} \big( \alpha_l'(1 -
\delta)^{+} + \sum_{i=1}^{L_T} (1-\alpha_l')(1 - \beta_i)^{+} < r
\big) \doteq \iint_{\mathcal{B}} p (\mathbb{\delta, \beta})
\textrm{d} \delta \textrm{d} \mathbb{\beta},\label{eqn:outage_24}
\end{IEEEeqnarray}
where $-\delta$ is the exponential order of $\sigma_{SD}^2$,
$-\beta_i$ is the exponential order of $\gamma(i)$, $\mathcal{B} =
\{\mathbb{\delta, \beta}: \alpha_l' (1-\delta)^{+} +
(1-\alpha_l')\sum_{i}(1 - \beta_i)^{+} < r\}$, and $p
(\mathbb{\delta, \beta})$ is the joint probability density function
of $\delta$ and $\beta$.
Substituting
\eqref{eqn:outage_22}-\eqref{eqn:outage_24} into
\eqref{eqn:outage_14}, the outage probability can be expressed as
\begin{eqnarray}
\textstyle\mathcal {P}_{out}^\star & \dot{\leq} &
\textstyle\rho_S^{-(M - K + 1 + N_r)(1 - r)^{+}} + \rho_S^{-(M - K +
1)(1 - \frac{r}{\alpha_l'})^{+}}\iint_{\mathcal{B}} p
(\mathbb{\delta, \beta}) \textrm{d} \delta \textrm{d}
\mathbb{\beta}, \label{eqn:outage_25}
\end{eqnarray}
and the DMT relation for the DF-MSC-opt scheme is given by the
following theorem.

\begin{Thm}
\label{Thm:DMT} The DMT relation for the DF-MSC-opt scheme can be
expressed as
\begin{eqnarray}
d(r,K)_{\textrm{DF-MSC-opt}} = \min ( d_1, d_2 + d_3 ),
\end{eqnarray}
where $d_1 = (M - K + 1 + N_r)(1 - r)^{+}$, $d_2 =(M - K + 1)(
\frac{1-2r}{1-r})^{+}$, and
\begin{eqnarray}
d_3 = \left\{
\begin{array}{l l}
\textstyle N_r + (N_r-1)K & 0 \leq r < 1/2\\
\textstyle \min \Big( \substack{N_r + (N_r - \theta)(K + 1 -
\theta)\\- (N_r + K - 2 \theta) (\frac{r}{1-r}-\theta)},
\substack{(N_r - \theta)(K + 1 - \theta)\\+N_r \theta
(\frac{1-r}{r})} \Big) & \frac{\theta}{\theta+1} \leq r <
\frac{\theta+1}{\theta+2} \\
\textstyle N_r L_T(\frac{1-r}{r}) & \frac{L_T}{L_T+1} \leq r \leq 1
\end{array}
\right.
\end{eqnarray}
with $\theta = 1,2,\ldots,L_T$ and $L_T = \min(N_r,K+1)$.
\end{Thm}
\proof Refer to Appendix \ref{pf:thm_DMT} for the proof.
\endproof
We could further optimize the parameter $K$ in the DF-MSC-opt scheme
and the resulting DMT relation is given by
\begin{IEEEeqnarray*}{Rl}
d(r)_{\textrm{DF-MSC-opt}}^\star\!=\!\!\!\max_{K \in [1,2,\ldots,M]}
\!\!d(r,K)_{\textrm{DF-MSC-opt}}\!=\!\!\max\big(d(r,\lfloor K^\star
\rfloor)_{\textrm{DF-MSC-opt}},d(r, \lceil K^\star \rceil
)_{\textrm{DF-MSC-opt}}\big),
\end{IEEEeqnarray*}
where
\begin{eqnarray}
K^\star = \left\{
\begin{array}{l l}
\frac{(M+1)(2-4r+r^2) + N_r(2-3r+r^2)}{(N_r-1)(1-r)+r^2} & 0 \leq r
\leq 1/2 \\
\frac{ (M+1)(1-r)^2 - N_r(3r-r^2-(1-r)2\theta) - \theta(3\theta
(1-r) - 1 - r)}{N_r(1-r) - r + (1-r)^2} & \frac{\theta}{\theta+1} <
r \leq \frac{\theta+1}{\theta+2}
\end{array}\label{eqn:KEddy}
\right.
\end{eqnarray}
is the optimal number of successfully decoding relay nodes to wait
for in the listening phase.

\begin{figure}[h]
\centering
\includegraphics[width = 3.5in]{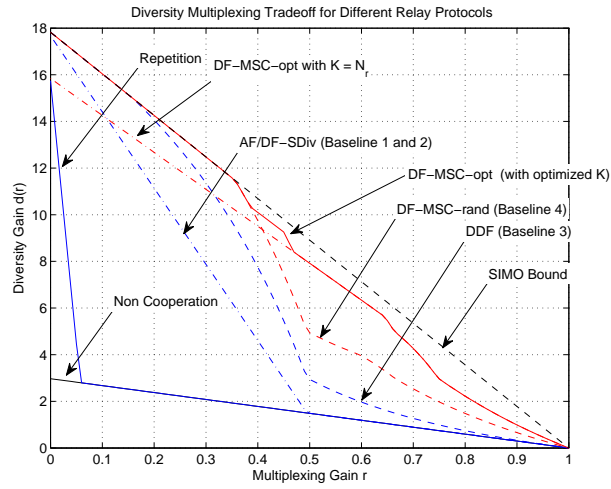}
\caption{Diversity-multiplexing tradeoff comparison of different
relay protocols.} \label{fig:dmt}
\end{figure}

In the following, we show that the DF-MSC-opt scheme achieves
superior DMT performance than the traditional cooperative diversity
schemes. Specifically, as per \cite{Tajer07}, the AF-SDiv and
DF-SDiv schemes have identical DMT relations given by
$d(r)_{\textrm{AF/DF-SDiv}} = M(1-2r)^{+} + N_r (1-r)^{+}$ for $0
\leq r \leq 1$. For the DDF
protocol, the DMT relation is given by the following lemma.
\begin{Lem}
\label{Lem:DDF} The DMT relation for the DDF protocol with $N_r$
receive antennas at the destination node can be expressed as
\begin{eqnarray}
d(r)_{\textrm{DDF}} = \left\{
\begin{array}{l l}
(M + N_r)(1-r) & 0 \leq r \leq \frac{N_r}{M + N_r} \\
N_r + M(\frac{1-2r}{1-r})^{+} & \frac{N_r}{M + N_r} \leq r \leq \frac{1}{2} \\
N_r(\frac{1-r}{r}) & \frac{1}{2} \leq r \leq 1
\end{array}
\right.
\end{eqnarray}
\end{Lem}
\proof Refer to Appendix \ref{pf:Lem:DDF} for the proof.
\endproof
In Fig.~\ref{fig:dmt}, we compare the DMT relations for the
DF-MSC-opt scheme and the baseline schemes. For a system with $M =
15$ relay nodes and a destination node with $N_r = 3$ antennas.
Note that since the DF-MSC-opt
scheme (as well as the DF-MSC-rand scheme) exploits multi-stream
cooperation, it can achieve high diversity gain than the traditional
cooperative diversity schemes (i.e. AF-SDiv, DF-SDiv, and DDF) that
exploit single-stream cooperation. Moreover, since the DF-MSC-opt
scheme optimizes the node selection policy and the number of
decoding relay nodes $K$, it can achieve better diversity gain than
the DF-MSC-rand scheme.

\subsection{TRT Analysis}
The DMT analysis alone cannot completely characterize the
fundamental tradeoff relation in the high SNR regime. Specifically,
the multiplexing gain gives the asymptotic growth rate of the
transmission rate $R$ at high SNR $\rho$ and is only applicable to
scenarios where $R$ scales {\em linearly} with $\log \rho$. Hence,
there are many unique transmission rates that correspond to the same
multiplexing gain, and the DMT analysis only gives a first order
comparison of the performance tradeoff at high SNR when we have
different multiplexing gains.

In order to have a clearer picture
on the tradeoff relations, in the following we study the SNR shift
in the outage probability $\mathcal P_{out}$ as we increase the
transmission rate $R$ by $\Delta R$. We quantify the more detailed
relations among the three parameters $(R, \log \rho, \mathcal
P_{out}(R, \rho))$ by analyzing the TRT\footnote{TRT analysis allows
for investigating more general scenarios where the transmission rate
$R$ does not scale linearly with $\log \rho$ and helps to study the
SNR gain of the outage probability vs SNR curve when we increase the
transmission rate $R$ by $\Delta R$. } relation \cite{Gamal07}.

Consider the outage probability expression for the DF-MSC-opt scheme
\eqref{eqn:outage_14}. In order to facilitate studying the TRT, we
choose the parameter $K$ such that the outage events are dominated
by relay-assisted transmissions, i.e. $K \geq \lceil K^\star \rceil$
(where $K^\star$ is given by \eqref{eqn:KEddy}). The following
theorem characterizes the asymptotic relationships among $R$,
$\rho_S$, and $\mathcal {P}_{out}^\star$ for the $r > \frac{1}{2}$
case\footnote{For the $r \leq \frac{1}{2}$ case, the MIMO channel
formed by the multiple relay nodes to the destination node can only
operate with a multiplexing gain of 1 and the TRT relation is
trivial.}.
\begin{Thm}
\label{Thm:3}  The TRT relation for the DF-MSC-opt scheme under $K
\geq \lceil K^\star \rceil $ and $r > \frac{1}{2}$ is given by
\begin{eqnarray}
\label{eqn:trt_1} \lim_{\rho_S \rightarrow \infty, (R,\rho_S) \in
\mathcal R(z)} \textstyle\frac{\log \mathcal {P}_{out}^\star -
c_{\textrm{DF-MSC-opt}}(z)R}{\log \rho_S} = -
g_{\textrm{DF-MSC-opt}}(z)
\end{eqnarray}
where
\begin{eqnarray}
\textstyle\mathcal R(z) \triangleq \textstyle\big\{(R, \rho_S)| z+1
> \frac{R}{(1-r)\log \rho_S} > z\big\}  \ \textrm{for} \ z \in
\mathbb Z, 0 \leq z < L_T,
\end{eqnarray}
denotes the $z^{\textrm{th}}$ operating region,
$c_{\textrm{DF-MSC-opt}}(z) \triangleq K + 1 + N_r - (2z + 1)$, and
$g_{\textrm{DF-MSC-opt}}(z) \triangleq (K + 1)N_r - z(z+1)$. Note
that $g_{\textrm{DF-MSC-opt}}(z)$ is defined as the reliability gain
coefficient and $t_{\textrm{DF-MSC-opt}}(z) \triangleq
g_{\textrm{DF-MSC-opt}}(z)/c_{\textrm{DF-MSC-opt}}(z)$ is defined as
the throughput gain coefficient.
\end{Thm}
\proof Refer to Appendix \ref{pf:Thm3} for the proof.
\endproof

\begin{figure}[h]
\centering
\includegraphics[width = 3.5in]{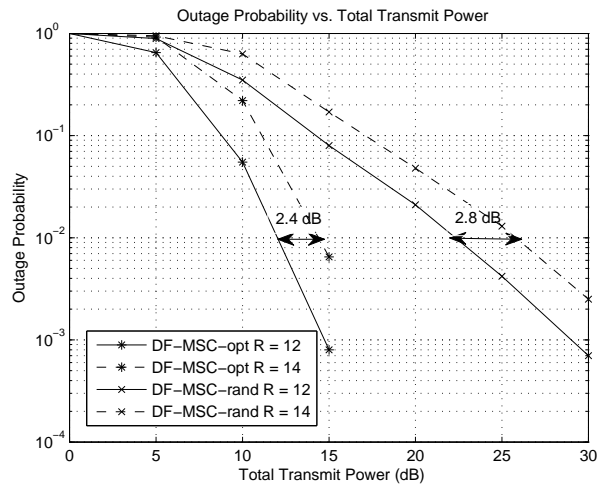}
\caption{Monte-Carlo simulation results for outage curves
corresponding to $\Delta$R = 2 bits/channel use for $N_r = 3$, $K =
3$ and $M = 15$ case with normalized S-R, R-D, S-D channel
variances.} \label{fig:trt_2}
\end{figure}

By applying Theorem \ref{Thm:3}, the SNR shift between two outage
curves with a $\Delta R$ rate difference is $3\Delta R/t(z)$ dB.
Fig.~\ref{fig:trt_2} shows the outage curves corresponding to
$\Delta R = 2$~bits/channel~use for $N_r = 3$, $K = 3$ and $M = 15$.
This scenario corresponds to the region $\mathcal R(1)$ and the TRT
relation for the DF-MSC-opt scheme can be expressed as
$g_{\textrm{DF-MSC-opt}}(1) = 10$ and $t_{\textrm{DF-MSC-opt}}(1) =
\frac{5}{2}$. As we can see from the simulation results, the SNR
shift is $2.4$ dB for 2
bits/channel use increase in the transmission rate, which matches
with our analysis.

\section{Conclusion}
\label{sect:conclusion} In this paper, we proposed a multi-stream
cooperative scheme (DF-MSC-opt) for multi-relay network. Optimal
multi-relay selection is considered and we derived the associated
outage capacity as well as the DMT and TRT relations. The proposed
design has significant gains in both the outage capacity as well as
the DMT relation due to (1) multi-stream transmissions in the
cooperative phase and (2) optimized relay selection.

\appendices

\section{Proof of Lemma \ref{Lem:exp}}
\label{pf:Lem_exp} To find the outage probability given $l$, we
characterize the function of $g(\mathbf H, \mathbf V^\star)$ under
different conditions.

\begin{itemize}
\item{ \bf Condition 1 $K < N_r$:} Under this condition, we allow all the successfully decoding relays to
transmit in the cooperative phase. Thus, the communication links can
be regarded as a conditional MIMO link \cite{Zheng03} and $g(\mathbf
H, \mathbf V) = \log\det\big(\mathbf I_{N_r} + \rho_S\mathbf
H_{D}(\mathcal D) \mathbf H_{D}(\mathcal D)^{H}\big) \leq \sum_{i =
1}^{K+1} \log_2(1 + \rho_S \sigma_{D}^2 \kappa(i))$, where
$\kappa(i)$, $i=1,\ldots,K+1$, are $\chi^2$-distributed variables
with $2N_r$ degrees of freedom.
\item{ \bf Condition 2 $K \geq N_r$:} Under this condition, we select $N_r$
nodes out of the source and the successfully decoding relay nodes
from the node selection space $\mathbf \Omega$ to participate in
cooperative transmission. The analytical solution for the term
$g(\mathbf H, \mathbf V)$ is in general not trivial
\cite{Molisch05}. Since in the proposed DF-MSC-opt scheme we choose
the \emph{best} $N_r$ out of $K+1$ transmit nodes, the upper bound
can be obtained similar to \cite[(6)]{Molisch05}. Thus, the capacity
bound with transmit node selection is $g(\mathbf H, \mathbf V) =
\max_{\mathbf V \in \mathbf \Omega}\log\det\big(\mathbf I_{N_r} +
\rho_S\mathbf H_{D}(\mathcal D) \mathbf V \mathbf V^{H} \mathbf
H_{D}(\mathcal D)^{H}\big) \leq \sum_{i = 1}^{N_r} \log_2(1 +
\rho_S\sigma_{D}^2 \kappa(i))$, where the $\kappa(i)$,
$i=1,\ldots,N_r$, are ordered $\chi^2$-distributed variables with
$2N_r$ degrees of freedom.
\end{itemize}

Combining the two conditions above, we obtain the general form as
shown in Lemma \ref{Lem:exp}.

\section{Proof of Theorem \ref{Thm:2}}
\label{pf:Thm:2} From \eqref{eqn:outage_14}, the only unknown part
is given by
\begin{eqnarray*}
\textstyle\textrm{Pr}\big(f(\alpha_l',\mathbf H, \mathbf V)<R\big) =
\textrm{Pr}\big[ \alpha_l' \log(1 + \rho_S|\mathbf H_{SD}|^2) + (1 -
\alpha_l')\sum_{i = 1}^{L_T} \log_2(1 + \rho_S \sigma_{D}^2
\kappa(i)) < R\big],
\end{eqnarray*}
which can be relaxed as
\begin{IEEEeqnarray*}{l}
\textstyle\textrm{Pr}\big[ \alpha_l' \log(1 + \rho_S|\mathbf
H_{SD}|^2) + (1 - \alpha_l')\sum_{i = 1}^{L_T} \log_2(1 + \rho_S
\sigma_{D}^2
\kappa(i)) < R\big]\\
\textstyle\leq \textrm{Pr}(\alpha_l' \log(1 + \rho_S| \mathbf
H_{SD}|^2) < R) \textrm{Pr}\big((1-\alpha_l')\sum_{i =
1}^{N_r} \log_2(1 + \rho_S \sigma_{D}^2 \kappa(i)) < R \big)\\
\textstyle= F\Big(\frac{2^{\frac{R}{\alpha_l'}} - 1 }{\rho_S
\sigma_{D}^2 };N_r\Big) \textrm{Pr}\Big(\sum_{i = 1}^{L_T}
\log_2\big(1 + \rho_S \sigma_{D}^2 \kappa(i) \big) <
\frac{R}{1-\alpha_l'}\Big).
\end{IEEEeqnarray*}
Moreover, the following relation holds for $\alpha_l'>0$
\begin{IEEEeqnarray}{l}
\textstyle\textrm{Pr}\Big(\sum_{i = 1}^{L_T} \log_2\big(1 + \rho_S
\sigma_{D}^2 \kappa(i)
\big) < \frac{R}{1-\alpha_l'}\Big) \leq \textrm{Pr}\Big(\log_2\big(1 + \rho_S \sigma_{D}^2 \kappa(1) \big) < \frac{R}{1-\alpha_l'}\Big) \nonumber \\
\;\;\;\;\;\;\;\;\;\;\;\;= \textstyle\textrm{Pr}\Big(\kappa(1) <
\frac{2^{\frac{R}{1-\alpha_l'}} - 1}{\rho_S \sigma_{D}^2}\Big) =
F\Big(\frac{2^{\frac{R}{1-\alpha_l'}} - 1 }{\rho_S \sigma_{D}^2
};N_r\Big)^{K + 1} \label{eqn:outage_15}
\end{IEEEeqnarray}
where in the last step we applied the results of order statistics
\cite{Ahsanullah01}. Substituting \eqref{eqn:outage_15} into
\eqref{eqn:outage_14}, we have Theorem \ref{Thm:2}.

\section{Proof of Theorem \ref{Thm:DMT}}
\label{pf:thm_DMT} The asymptotic expression of the outage
probability $\mathcal {P}_{out}^\star$ is given by
\eqref{eqn:outage_25} with $\mathcal{B} = \{\mathbb{\delta, \beta}:
\alpha_l' (1-\delta)^{+} + (1-\alpha_l')\sum_{i}(1 - \beta_i)^{+} <
r\}$.

From the first term in \eqref{eqn:outage_25}, we can obtain $d_1 =
(M-K+1+N_r)(1-r)^{+}$.

Since the source node is equipped with a single antenna, the maximum
multiplexing gains for the S-R links as well as the S-D and R-D
links is always less than 1. Specifically, the multiplexing gain in
the listening phase is $\frac{r}{\alpha_l'} \leq 1$ and the
multiplexing gain in the cooperative phase is $\frac{r}{1 -
\alpha_l'} \leq 1$ or equivalently, $\alpha_l' \geq \max(r,1-r)$.
Thus, we can evaluate the second term through $d_2 =
(M-K+1)(1-\frac{r}{1-r})$ for $r < 1/2$ and $0$ otherwise.

We evaluate the third term under the following cases. When $r<1/2$,
$\alpha_l' \geq 1-r$ and $\mathcal{B} \subseteq \{\mathbb{\delta,
\beta}: r \big( (1-\delta)^{+} + \sum_{i}(1 - \beta_i)^{+} \big) <
r\}$. Equivalently,
\begin{eqnarray}
& \textstyle\iint_{\mathcal{B}} p (\mathbb{\delta, \beta})
\textrm{d} \delta \textrm{d} \mathbb{\beta} & \leq \textstyle
\iint_{(1-\delta)^{+} + \sum_{i}(1 - \beta_i)^{+}\leq 1}p
(\mathbb{\delta, \beta})
\textrm{d} \delta \textrm{d} \mathbb{\beta} \nonumber \\
& & \textstyle\doteq \rho_S^{-\inf_{r' \in [0,1]}\big( N_r(K+1) -
(N_r + K)r' + N_r r'\big)} = \rho_S^{-(N_r-1)K - N_r}.
\end{eqnarray}
As a result, $d_3 = (N_r-1)K + N_r$ for $r<1/2$. When $1/2 \leq
r<1$, $\alpha_l' \geq r$ and the corresponding $\mathcal{B} =
\{\mathbb{\delta, \beta}:  r (1-\delta)^{+} + (1-r) \sum_{i}(1 -
\beta_i)^{+} r < r\}$. Equivalently,
\begin{eqnarray}
&\textstyle\iint_{\mathcal{B}} p (\mathbb{\delta, \beta}) \textrm{d}
\delta \textrm{d} \mathbb{\beta} &\textstyle\leq
\iint_{(1-\delta)^{+} + \frac{1-r}{r}\sum_{i}(1 - \beta_i)^{+}\leq
1}p (\mathbb{\delta,
\beta}) \textrm{d} \delta \textrm{d} \mathbb{\beta} \nonumber \\
& &\textstyle\doteq \rho_S^{-\inf_{r' \in [0,1]} \big(
(N_r-\frac{rr'}{1-r})(K+1-\frac{rr'}{1-r}) - N_r r' \big)}
\end{eqnarray}
for $\frac{rr'}{1-r} \in \{1,\ldots,L_T\}$. Thus, we can write the
general form as $d_3 = \inf_{0 \leq r' \leq 1} \big( N_r r' + (N_r -
\theta)(K + 1 - \theta) - (N_r + K - 2 \theta)
(\frac{rr'}{1-r}-\theta) \big)$ for $\theta = \{1,2,\ldots,L_T\}$.
Since $d_3$ is linear with respect to $r'$, we can write $d_3 = \min
\big\{ N_r + (N_r - \theta)(K + 1 - \theta) - (N_r + K - 2 \theta)
(\frac{r}{1-r}-\theta), N_r \theta (\frac{1-r}{r}) + (N_r -
\theta)(K + 1 - \theta) \big\}$. Moreover, when $\frac{r}{1-r} \geq
L_T$, we have $\frac{rr'}{1-r}-L_T = 0$ and $d_3 = N_r
L_T(\frac{1-r}{r})$.

Combining the three terms above, we have Theorem \ref{Thm:DMT}.

\section{Proof of Lemma \ref{Lem:DDF}}
\label{pf:Lem:DDF} The main steps of the proof are based on the work
by \cite{Azarian05}. Denote $g_{k,j}$ to be the channel coefficient
between the $j^{th}$ node and the $k^{th}$ receive antenna at the
destination node, and the received signal at the $k^{th}$ antenna is
given by $y_k = g_{k,j}x_j + z_k$. Following the same argument in
the proof of Theorem 6 in \cite{Azarian05}, we can find that
destination¡¯s ML error probability assuming {\em error-free}
decoding at all of the relays, provides a lower bound on the
diversity gain achieved by the protocol and the corresponding PEP
for $k^{th}$ antenna is upper bounded by $P_k \leq \prod_{j=1}^{M+1}
\left[1 + \big(\sum_{i=1}^j |g_{k,i}|^2\big)\rho \right]^{-n_j}$,
where $n_j$ is the number of symbol intervals in the codeword during
which a total of $j$ nodes are transmitting, so that, the total
number of codeword length $\sum_{j=1}^{M+1} n_j = N$. Hence, the PEP
for the destination to received the information is upper bounded by
$P \leq \prod_{k=1}^{N_r} P_k$. Note that, the upper bound is not
tight since we do not consider the joint decoding among the receive
antennas, but it is sufficient to derive the order-wise relation. By
changing the order of $k$ and $j$, we have $P \leq \prod_{j=1}^{M+1}
\left[\prod_{k=1}^{N_r}\Big(1 + \big(\sum_{i=1}^j
|g_{k,i}|^2\big)\rho\Big) \right]^{-n_j}$. $\sum_{j=1}^{p} n_j$ is
the number of symbol intervals that relay $p$ has to wait, before
the mutual information between its received signal and the signals
that the source and other relays transmit exceeds $NR$. Thus, we
have $\sum_{j=1}^{p} n_j \leq \min \left\{N, \frac{NR}{\log(1+
|h_{p+1,1}|^2 \rho)}\right\}$, where $h_{p+1,1}$ denotes the channel
condition between the relay node $p$ and the source node.

Define $v_{k,j}$ and $u_{j,i}$ as the exponential orders of
$g_{k,i}$ and $h_{j,i}$. We have the following relation,
\begin{IEEEeqnarray*}{l}
P \leq \rho^{-\sum_{j=1}^M n_j\left(1 -
\min\{\sum_{k=1}^{N_r}v_{k,1}, \ldots,
\sum_{k=1}^{N_r}v_{k,j}\}\right)^{+}} = \rho^{-\{\sum_{j=1}^M
n_j\left(1 - N_r \min \{v_1, \ldots, v_j\}\right)^{+}},
\end{IEEEeqnarray*}
where $v_i = \frac{\sum_{k=1}^{N_r}v_{k,i}}{N_r}$. Choose the rate
$R = r \log \rho$ and the PEP bound for rate $R$ is given by
\begin{IEEEeqnarray*}{l}
P \leq \rho^{-\sum_{j=1}^M n_j\left(1 -
\min\{\sum_{k=1}^{N_r}v_{k,1}, \ldots,
\sum_{k=1}^{N_r}v_{k,j}\}\right)^{+}} = \rho^{-N
\left[\{\sum_{j=1}^M \frac{n_j}{N}\left(1 - N_r \min \{v_1, \ldots,
v_j\}\right)^{+}- r \right]}.
\end{IEEEeqnarray*}
Hence, the set of channel realizations that satisfy $\{\sum_{j=1}^M
\frac{n_j}{N}\left(1 - N_r \min \{v_1, \ldots, v_j\}\right)^{+} \leq
r$ results in the outage event.

Define $\bar{v_j} = \min\{v_1, \ldots, v_j\}$ and we can separate
the discussion in the following three cases. (Naturally, we have
$\bar{v_1} \geq \bar{v_2} \geq \ldots \geq \bar{v_M}$).
\begin{itemize}
\item Case 1: $\bar{v_1} \leq 1$. In this case, we can follow the discussion from (62) to (71) in \cite{Azarian05} and show that the diversity order $d$ is $d \geq \left\{
    \begin{array}{c c}
    (N_r+M)(1-r), & \frac{N_r}{N_r+M} \geq r \geq 0 \\
    N_r + M (\frac{1-2r}{1-r}), & \frac{1}{2} \geq r > \frac{N_r}{N_r+M} \\
    N_r(\frac{1-r}{r}), & 1 \geq r > \frac{1}{2}
    \end{array}
    \right.$.
\item Case 2: $\bar{v_M} \geq 1$. It's trivial and $d = M$.
\item Case 3: $\bar{v_i} > 1 \geq \bar{v_{i+1}}$. Follow the same discussion from (72) to (82) in \cite{Azarian05} , we conclude that $d \geq \left\{
    \begin{array}{c c}
    N_r + M (\frac{1-2r}{1-r}), & \frac{1}{2} \geq r \geq 0 \\
    N_r(\frac{1-r}{r}), & 1 \geq r > \frac{1}{2}
    \end{array}
    \right.$.
\end{itemize}

Combining the above cases, we have Lemma \ref{Lem:DDF}.

\section{Proof of Theorem \ref{Thm:3}}
\label{pf:Thm3} Given $r > \frac{1}{2}$ and $K \geq \lceil
K^\star\rceil$, \eqref{eqn:outage_14} can be simplified as $\mathcal
{P}_{out}^\star \dot{\leq} - \Phi'_{l'}$ $\textrm{Pr}(
f(\alpha_l',\mathbf H, \mathbf V^\star) < R)$. Substituting the
expression of $f(\alpha_l',\mathbf H, \mathbf V)$, we have
\begin{eqnarray}
\label{eqn:pf_Thm_2_0} \textstyle\mathcal {P}_{out}^\star & \leq &
\textstyle \textrm{Pr}( \alpha_l' \log(1 + \rho_S|\mathbf H_{SD}|^2)
+ (1 - \alpha_l')\sum_{i = 1}^{L_T} \log_2(1 + \rho_S \sigma_{D}^2
\gamma(i) ) < R) \nonumber \\
& \leq & \textstyle \textrm{Pr}( (1 - r) \sum_{i = 1}^{L_T} \log_2(1
+ \rho_S \sigma_{D}^2 \gamma(i) ) < R).
\end{eqnarray}
Based on the above results and following the same lines as the proof
of \cite[Theorem 2]{Gamal07}, we have the result of Theorem
\ref{Thm:3}.

\bibliographystyle{IEEEtran}
\bibliography{IEEEabrv,mybibfile}

\begin{IEEEbiography}[{\includegraphics[width=1in,height=1.25in,clip,keepaspectratio]{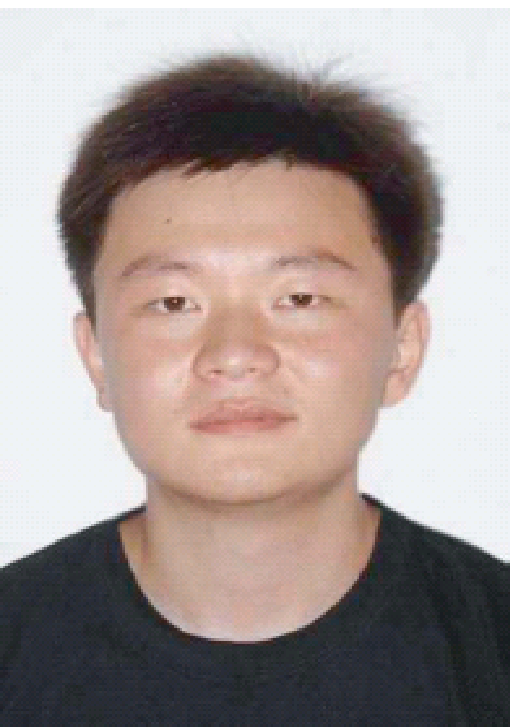}}]{Dr. Shunqing Zhang} obtained B.Eng from Fudan University and Ph.D.
from Hong Kong University of Science and Technology (HKUST) in 2005
and 2009, respectively. He joined Huawei Technologies Co., Ltd in
2009, where he is now the system engineer of Green Radio Excellence
in Architecture and Technology (GREAT) team. His current research
interests include the energy consumption modeling of the wireless
system, the energy efficient wireless transmissions as well as the
energy efficient network architecture and protocol design.
\end{IEEEbiography}

\begin{IEEEbiography}[{\includegraphics[width=1in,height=1.25in,clip,keepaspectratio]{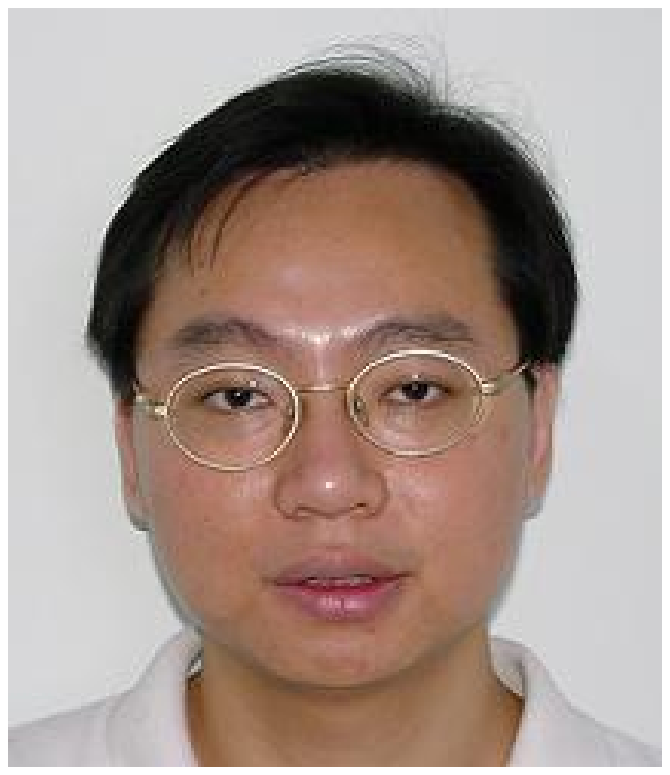}}]{Vincent
K. N. Lau} obtained B.Eng (Distinction 1st Hons) from the University
of Hong Kong in 1992 and Ph.D. from Cambridge University in 1997. He
was with PCCW as system engineer from 1992-1995 and Bell Labs -
Lucent Technologies as member of technical staff from 1997-2003. He
then joined the Department of Electronic and Computer Engineering,
Hong Kong University of Science and Technology as Professor. His
current research interests include robust and delay-sensitive
cross-layer scheduling of MIMO/OFDM wireless systems with imperfect
channel state information, cooperative and cognitive communications
as well as stochastic approximation and Markov Decision Process.
\end{IEEEbiography}

\vfill

\newpage
\begin{table}[!h]
\renewcommand{\arraystretch}{1.3}
\caption{Comparisons among the DF-MSC-opt scheme and the baseline
schemes.} \label{table:DF-MSCvsBaselines} \centering
\begin{tabular}{|c|c|c|c|c|c|}
\hline
\parbox{3.75cm}{\textbf{Relay protocol}} & \textbf{AF-SDiv} & \textbf{DF-SDiv} & \textbf{DDF} & \textbf{DF-MSC-rand} & \textbf{DF-MSC-opt}\\
\hline
\parbox{3.75cm}{\textbf{No. of transmitted streams in the cooperative phase}} & 1 & 1 & 1 & $N_r$ & $N_r$\\
\hline
\parbox{3.75cm}{\textbf{No. of relay nodes in the cooperative phase}} & $M$ &
\parbox{1.5cm}{depends on\\S-R links} & \parbox{1.5cm}{depends on\\S-R links} & $N_r$ & \parbox{1.4cm}{$N_r$ chosen from $K$}\\
\hline
\parbox{3.75cm}{\textbf{Receiver structure}} & MRC & MRC & MRC & ML & ML\\
\hline
\end{tabular}
\end{table}

\newpage

\end{document}